\documentclass[reprint,amsmath,amssymb,aps,pra,showpacs,floatfix]{revtex4-1}
\usepackage{graphicx}
\pdfoutput=1

\begin{document}

\title{Correlation control for pure and efficiently generated heralded single photons}

\author{Jefferson Fl\'orez}
\email[]{j.florez34@uniandes.edu.co}
\affiliation{Laboratorio de \'Optica Cu\'antica, Universidad de los Andes, A.A. 4976, Bogot\'a D.C., Colombia}

\author{Omar Calder\'on}
\affiliation{Laboratorio de \'Optica Cu\'antica, Universidad de los Andes, A.A. 4976, Bogot\'a D.C., Colombia}

\author{Clara I. Osorio}
\affiliation{FOM Institute AMOLF, Science Park 104, 1098 XG, Amsterdam, The Netherlands}

\author{Alejandra Valencia}
\affiliation{Laboratorio de \'Optica Cu\'antica, Universidad de los Andes, A.A. 4976, Bogot\'a D.C., Colombia}%

\begin{abstract}
We present a detailed study on the properties of single photons generated by spontaneous parametric down conversion (SPDC) when both the spectral and spatial degrees of freedom are controlled by means of filters. Our results show that it is possible to obtain pure heralded single photons with high heralding efficiency despite the use of filters. Moreover, we report an asymmetry on the single photon properties exhibited in type-II SPDC sources that depends on choosing the signal or the idler photon as the heralding one.
\end{abstract}


\maketitle

\section{Introduction} \label{Sec:Intro}

Spontaneous parametric down conversion (SPDC) is a convenient source of paired photons \cite{Shih03}. Due to the intrinsic properties of the non-linear process that leads to pair generation, SPDC photons are correlated in space, frequency and other degrees of freedom. Detailed studies of these correlations have been carried out to understand the SPDC process itself \cite{Rubin96} and its promising use for quantum applications \cite{Scarcelli03,Giovannetti01}. For example, studies of the spatial and spectral correlations have been widely investigated due to the role of SPDC on the generation of what is called heralded single photons (HSPs) \cite{Eisaman11}. In a HSP source, the detection of one of the photons announces the presence of its partner, which is then available for applications.

For most applications, from a single photon source it is desirable that all of the photons generated are pure, indistinguishable and produced on-demand \cite{Eisaman11}. In order to get a pure HSP based on SPDC, it is important to remove the spatial and spectral correlations within the photon pair. There are various approaches to remove the spectral correlations by engineering the frequency joint-spectrum of the pairs \cite{Valencia07}, using achromatic phase matching \cite{Torres05,Hendrych07}, or working with the appropriate crystal at the appropriate wavelength \cite{Mosley08}. On the other hand, regarding the engineering of spatial SPDC correlations, some studies show how different pump spectral profiles result in different spatial correlations \cite{Hamar10,Molina-Terriza05}.

Spatial and spectral filtering is the most common way to control and even suppress the correlations of SPDC photons \cite{Osorio08,Aichele02}. Filters have the advantage of being relatively simple to implement experimentally, but with the drawback of decreasing the amount of photons available. However, in recent years, intense sources of photon pairs have appeared \cite{Steinlechner12} opening the possibilities to control correlations by means of filters. Motivated by these recent advances, in this paper we describe the purity and heralding efficiency of HSP taking into account both spatial and spectral filtering.

Our study is based on the mathematical formalism developed in Ref. \cite{Osorio08}, which provides a compact tool to study the role of the SPDC parameters and the filtering process on the SPDC spatial and spectral correlations. From our study, we found that when using type-II SPDC to generate HSPs, it is different to announce with the signal or the idler. We show that for some experimental parameters, the heralding efficiency associated to each of the photons is different. This result is in agreement with reported experimental results as the ones in \cite{Steinlechner12}. In addition, we found that filtered SPDC sources allow the production of HSP with high heralding efficiency. This differs from the previous conception according to which the use of filters was thought to be detrimental for the heralding efficiencies, at least when both SPDC photons were filtered \cite{Mosley08}. The results reported in this paper make HSP based on SPDC a more promising source given the new high intense sources of paired photons.

This paper is organized as follows: Sec.~\ref{Sec:Theory} contains mathematical expressions for the SPDC two-photon state and HSP properties; the effects of filtering on the spatial HSP purity and heralding efficiency are discussed in Sec.~\ref{Sec:Spatial} in the presence and absence of the spatial-spectral correlation; in Sec.~\ref{Sec:Spectral}, we analyze the spectral HSP properties and identify the conditions for which it is possible to obtain a pure HSP with high heralding efficiency; finally, in Sec.~\ref{Sec:Conclusions} we present our conclusions.

%%%%%%%%%%%%%%%%%%%%%%%%%%%%%%%%%%%%%%%%%%%%%%%%%%%%%%%%%%%%%%%
\section{Theoretical framework} \label{Sec:Theory}

To understand the role of filters in the generation of HSPs via SPDC, in this section we develop theoretically the spatial and spectral correlations of paired photons. In addition, we define the purity and heralding efficiency of HSP and its relationship with the characteristics of the SPDC photon pairs.

\subsection{Spatial and spectral SPDC two-photon state}\label{subsec:PhotonPairState}

\begin{figure}[htb]
  \includegraphics[width=0.3\textwidth]{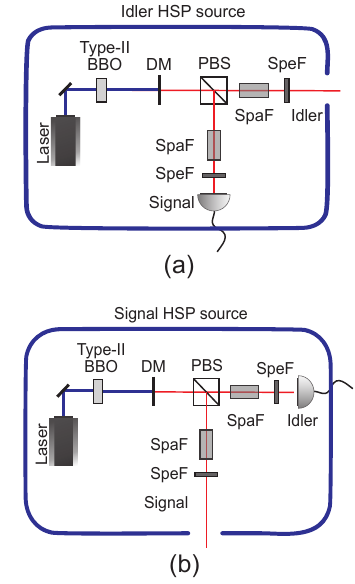}\\
  \caption{(Color online) Experimental setup for a HSP source using a type-II BBO crystal in a collinear configuration. A dichroic mirror (DM) is used for stoping the pump, and a polarizing beamsplitter (PBS) separates the down-converted photons according to their polarization. Spatial (SpaF) and spectral (SpeF) filters are placed in the heralding- and heralded-photon paths. In (a), the idler (ordinary-polarized photon) is the HSP, while in (b) the signal (extraordinary-polarized photon) is the HSP.}\label{fig:Setup}
\end{figure}

SPDC is a non-linear optical process in which photons from an intense pump beam are occasionally divided in two photons known as signal and idler. Fig. \ref{fig:Setup} illustrates a typical experimental setup for the generation of HSP via filtered SPDC. A pump laser impinges into a non-linear crystal where pairs of photons are produced spontaneously. Using first order perturbation theory, in the paraxial approximation, the two-photon state as a function of the frequencies $\omega_\mu$ ($\mu=s,i$) and transverse wavevectors $\mathbf{q}_\mu=(q_\mu^x,q_\mu^y)$ of the signal (s) and the idler (i) photons, is given by \cite{Rubin94}
\begin{multline}
|\psi\rangle=\int d^2\mathbf{q}_sd^2\mathbf{q}_i d\omega_s d\omega_i \phi(\mathbf{q}_s,\omega_s,\mathbf{q}_i,\omega_i)\\
\times\hat{a}_s^\dagger(\mathbf{q}_s,\omega_s)\hat{a}_i^\dagger(\mathbf{q}_i,\omega_i)|0\rangle,
\label{eq:PhotonPairState}
\end{multline}
where $\phi(\mathbf{q}_s,\omega_s,\mathbf{q}_i,\omega_i)$ is the so-called mode function or biphoton, which play the role of a joint-probability amplitude for the photon-pair state. This function contains all the information about the spatial and spectral correlations between the photons in the Hilbert space spanned by $\mathbf{q}_s$, $\mathbf{q}_i$, $\omega_s$, and $\omega_i$, and the role of the filters.

For a collinear SPDC process, occurring in a crystal of length $L$, the mode function is
\begin{multline}
\phi(\mathbf{q}_s,\omega_s,\mathbf{q}_i,\omega_i)=N\alpha(\mathbf{q}_s+\mathbf{q}_i)\beta(\omega_s+\omega_i)\text{sinc}\left(\frac{\Delta kL}{2}\right)\\ \times\exp\left(\frac{i\Delta kL}{2}\right)f_s(\omega_s)f_i(\omega_i)C_s(\mathbf{q_s})C_i(\mathbf{q_i}).
\label{eq:MF}
\end{multline}
Here, the spatial and spectral filters are denoted by $C_\mu(\mathbf{q}_\mu)$ and $f_\mu(\omega_\mu)$, respectively, $N$ is a normalization constant, $\alpha(\mathbf{q}_s+\mathbf{q}_i)$ and $\beta(\omega_s+\omega_i)$ describe the spatial and spectral pump distributions, respectively, and $\Delta k$ accounts for the longitudinal phase mismatching between the involved fields.

\begin{figure}[htb]
  \includegraphics[width=0.45\textwidth]{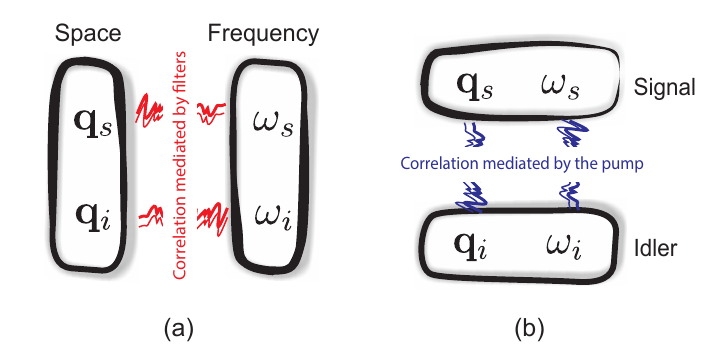}\\
  \caption{(Color online) Illustration of the correlations associated to the SPDC photon pair generation. In (a), the two-photon system is viewed as composed by the spatial and spectral variables of each photon. Along this work, the correlation between these two degrees of freedom is controlled by means of filters. In (b), the two-photon system is considered as one given by the signal variables and the other by the idler ones. The signal-idler correlation is mediated through the pump properties.}\label{fig:AllCorrelations}
\end{figure}

In general, the mode function in Eq. (\ref{eq:MF}) cannot be separated into two subsystems defined by neither $\{\mathbf{q}_s,\mathbf{q}_i\}$ and $\{\omega_s,\omega_i\}$, nor by $\{\mathbf{q}_s,\omega_s\}$ and $\{\mathbf{q}_i,\omega_i\}$, which means that the two-photon system exhibit correlations between space and frequency and between signal and idler \cite{Osorio08}. Fig. \ref{fig:AllCorrelations} shows a pictorial representation of such correlations. In this work, the correlation between space and frequency is controlled by means of spatial and spectral filters placed in the signal and idler paths, as indicated in Fig. \ref{fig:AllCorrelations}(a), while the degree of signal-idler correlation is mediated by the spatial and spectral properties of the pump, as in Fig. \ref{fig:AllCorrelations}(b).

%If spatial and spectral filters are placed in the signal and idler paths, as shown in Fig. \ref{fig:Setup}, the mode function becomes
%\begin{multline}
%\tilde{\phi}(\mathbf{q}_s,\omega_s,\mathbf{q}_i,\omega_i)=\phi(\mathbf{q}_s,\omega_s,\mathbf{q}_i,\omega_i)\\
%\times f_s(\omega_s)f_i(\omega_i)C_s(\mathbf{q_s})C_i(\mathbf{q_i}),
%\label{eq:FMF}
%\end{multline}
%where the spatial and spectral filters are denoted by $C_\mu(\mathbf{q}_\mu)$ and $f_\mu(\omega_\mu)$, respectively.

Following \cite{Osorio08}, under certain approximations it is possible to write Eq.~(\ref{eq:MF}) using a matrix representation such that
\begin{equation}
\phi(\mathbf{q}_s,\Omega_s,\mathbf{q}_i,\Omega_i)=N\exp{(x^T Ax)},
\label{matrix}
\end{equation}
with $A$ a $6\times 6$ matrix that depends on all the parameters of the SPDC process and the filters, $x$ a vector whose transpose is $x^T=(q_s^x,q_s^y,q_i^x,q_i^y,\Omega_s,\Omega_i)$, and $\Omega_\mu=\omega_\mu-\omega_\mu^0$ the frequency deviation from the central frequency, $\omega_\mu^0$. The necessary approximations to write Eq.~(\ref{matrix}) are not far from experimental realization and are the following: a) Gaussian distributions for the spatial and spectral profiles of the pump beam, $\alpha(\mathbf{q}_p)\propto\exp{\left[-\text{w}^2_p\mathbf{q}_p^2/4\right]}$ and $\beta(\Omega_p)\propto\exp{\left[-\Omega_p^2/4\sigma_p^2\right]}$, with $\text{w}_p$ the pump waist and $\sigma_p$ the pump spectral bandwidth; b) Gaussian filters characterized by a spectral bandwidth, $\sigma_\mu$, and a spatial collecting mode, $\text{w}_\mu$, in such a way that $f_\mu(\Omega_\mu)\propto\exp{\left[-\Omega_\mu^2/4\sigma_\mu^2\right]}$ and $C_\mu(\mathbf{q}_\mu)\propto\exp{\left[-\text{w}^2_\mu\mathbf{q}_\mu^2/4\right]}$; c) the sinc function in Eq.~(\ref{eq:MF}) is approximated by $\text{sinc(x)}\approx\exp{(-\gamma x^2)}$, where $\gamma$ equal $0.193$ makes both functions to have the same width at $1/e^2$ of their maxima; and d) $\Delta k$ is expanded up to first order in Taylor's series around a central frequency $\omega_\mu^0$, obtaining
\begin{equation}
\Delta k\approx(\rho_p-\rho_s)q_s^x+\rho_pq_i^x+D_s\Omega_s-D_i\Omega_i.
\label{Deltak}
\end{equation}
Here $D_\mu$ is the inverse group velocity difference between the photon $\mu$ and the pump, and $\rho_\epsilon$ ($\epsilon=p,s$) is the walk-off angle. This angle is given by $\rho_\epsilon=-\frac{1}{n_\epsilon}\frac{\partial n_\epsilon}{\partial \theta_\epsilon}$, where $n_\epsilon$ is the effective refractive index for the extraordinary beam and $\theta_\epsilon$ is the angle formed by the wavevector $\mathbf{k}_\epsilon$, and the optical axis of the crystal.

\begin{figure}
  \includegraphics[width=0.45\textwidth]{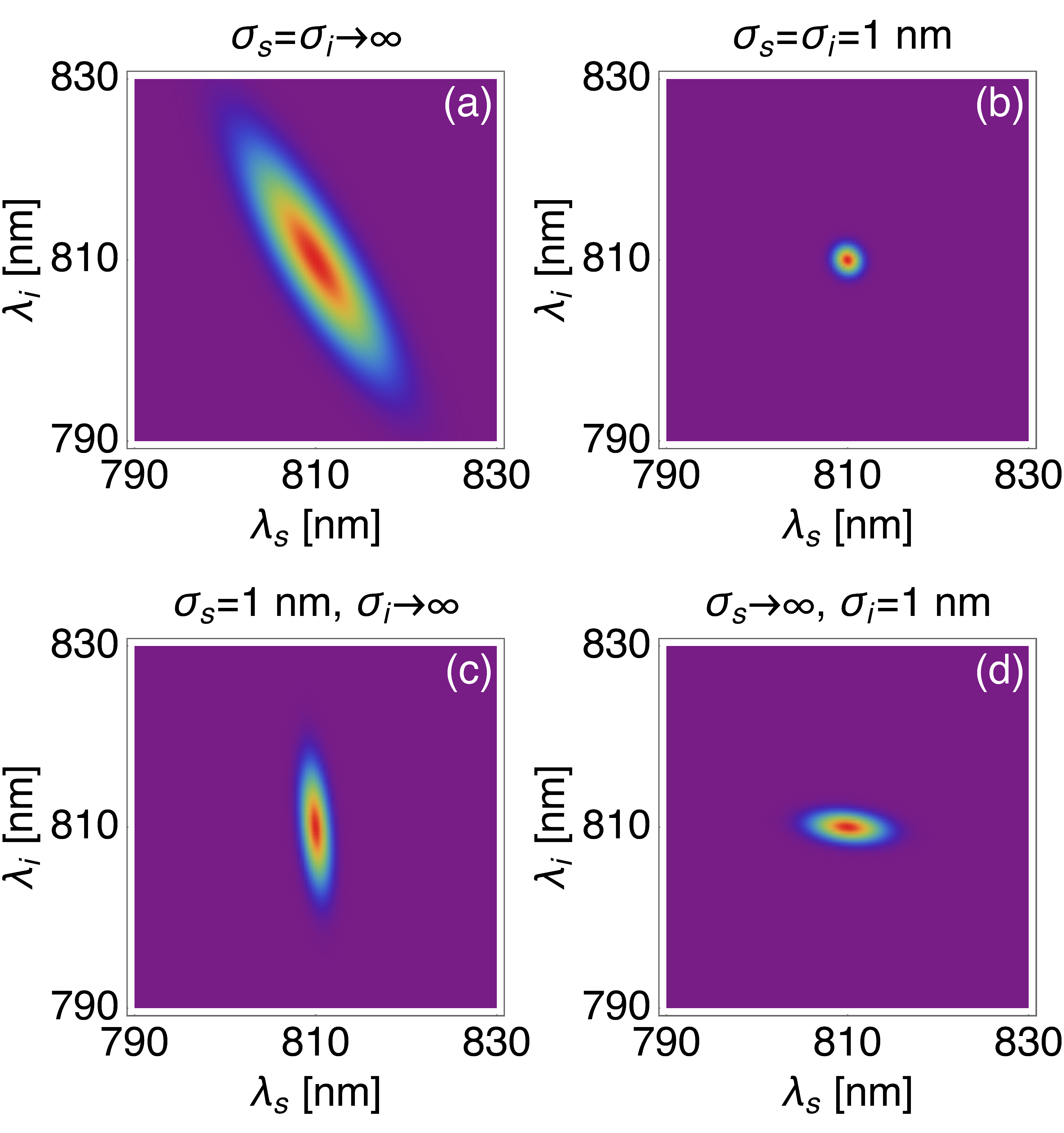}\\
  \caption{(Color online) Effect of filtering on the spectral joint-spectrum for collinear type-II SPDC. In (a) spectral filters are not considered, while in (b) both arms are filtered using the same filter bandwidth. In (c) and (d) only one of the photons is spectrally filtered.}
  \label{fig:effectofspectralfiltering}
\end{figure}

\begin{figure}
  \includegraphics[width=0.45\textwidth]{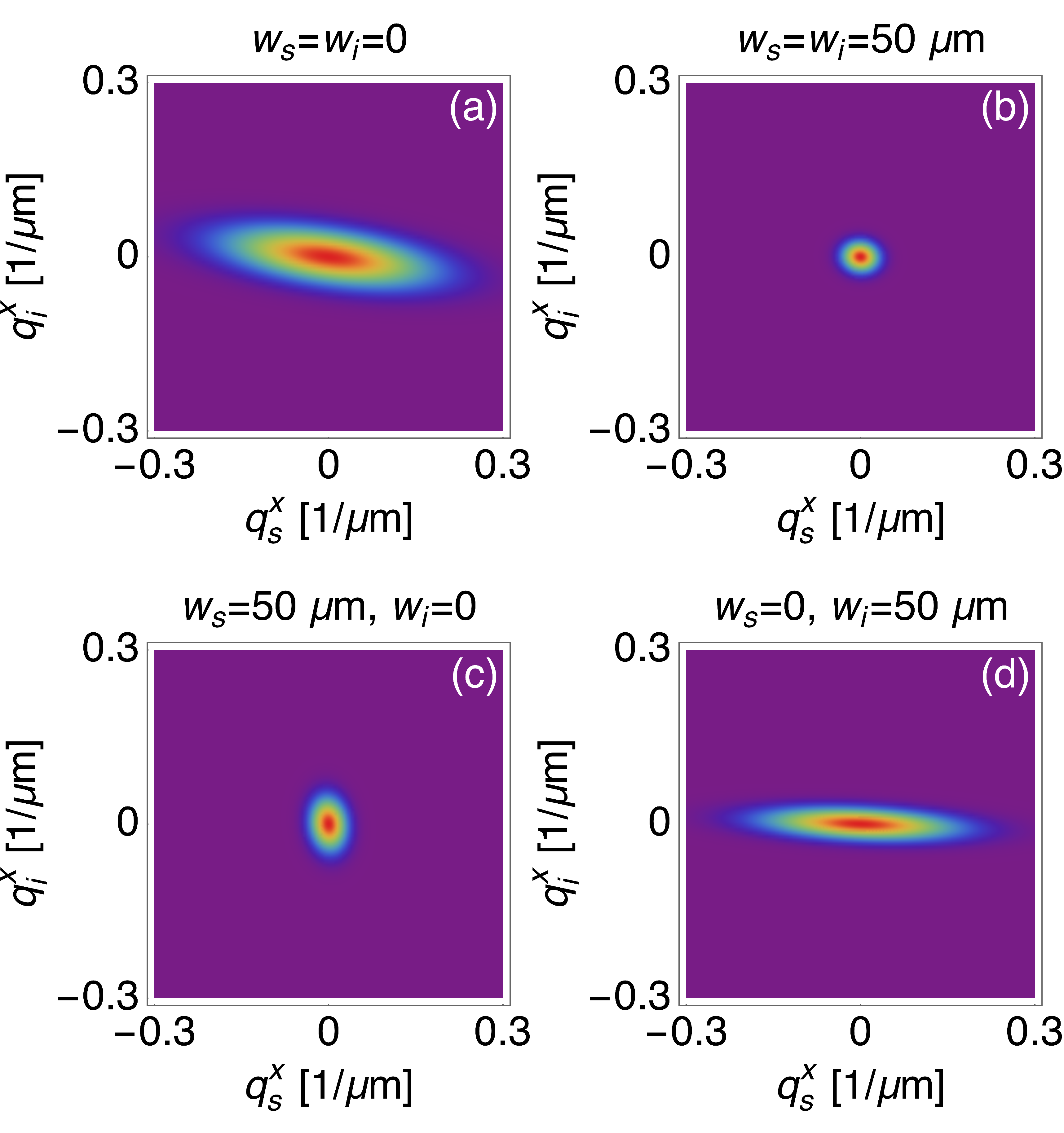}\\
   \caption{(Color online) Effect of filtering on the spatial joint-spectrum for collinear type-II SPDC. In (a) spatial filters are not considered, while in (b) both arms are filtered using the same collecting mode. In (c) and (d) only one of the photons is spatially filtered.}
   \label{fig:effectofspatialfiltering}
\end{figure}

An idea about the spectral and spatial correlations of the SPDC photons and the effect that filters have on it, can be obtained by looking at the joint-spectrum, $|\tilde{\phi}(\mathbf{q}_s,\Omega_s,\mathbf{q}_i,\Omega_i)|^2$. It is convenient to consider independently the spectral, $|\tilde{\phi}(\mathbf{q}_s=0,\Omega_s,\mathbf{q}_i=0,\Omega_i)|^2$, and the spatial, $|\tilde{\phi}(\mathbf{q}_s,\Omega_s=0,\mathbf{q}_i,\Omega_i=0)|^2$, joint-spectrum. Fig.~\ref{fig:effectofspectralfiltering}(a) and Fig.~\ref{fig:effectofspatialfiltering}(a) depict the corresponding unfiltered biphoton ($\sigma_s=\sigma_i\to\infty$ and $\text{w}_s=\text{w}_i=0$, respectively) for collinear type-II SPDC, produced in a $1$ mm BBO pumped by a laser centered at $405$ nm with $\sigma_p=1$ nm and $\text{w}_p=10$ $\mu$m. The tilt with respect to the straight line at $-45^{\circ}$ in Fig. \ref{fig:effectofspectralfiltering}(a) and Fig. \ref{fig:effectofspatialfiltering}(a) is due to the fact that we are considering a type II process in which signal and idler have orthogonal polarizations. The tilt in the spectral domain is a consequence of the different group velocities for signal and idler photons inside the nonlinear crystal. In the spatial domain, the observed tilt is due to the presence of the walk-off angle for the signal photons that is absence for the idler ones.

The presence of the tilt in the spectral and spatial domains has consequences at the moment of filtering the SPDC source. When  the same filter is placed in the signal and idler paths, Fig. \ref{fig:effectofspectralfiltering}(b) for the spectral regime and Fig. \ref{fig:effectofspatialfiltering}(b) for the spatial domain, the joint-spectrum looses the information of the tilt. However, when only one of the photons is filtered, the areas of the joint spectral and joint spatial functions are different, as can be seen comparing Fig. \ref{fig:effectofspectralfiltering}(c) with Fig. \ref{fig:effectofspectralfiltering}(d), and Fig. \ref{fig:effectofspatialfiltering}(c) with Fig. \ref{fig:effectofspatialfiltering}(d). This fact reveals that, for a type-II SPDC source, to place a filter in the signal arm leads to different results than a filter in the idler arm. 

To get further insight into the spatial and spectral domain of the SPDC, it is useful to work within the density matrix formalism. When spatial and spectral filters are considered in the paths of signal and idler photons, the density matrix is $\hat{\rho}=|\psi\rangle\langle \psi|$, with $|\psi\rangle$ the SPDC two-photon state given by Eq. (\ref{eq:MF}).

If one is interested in the spatial or spectral properties of the two-photon state, one must apply partial traces to $\hat{\rho}$ and define reduced density matrices. For example, the signal-idler spatial density matrix is
\begin{equation}
\label{ro-biphoton-spatial}
\hat{\rho}_{\mathbf{q}}=\text{Tr}_{\Omega}(\hat{\rho}),
\end{equation}
and the signal-idler spectral density matrix is
\begin{equation}
\label{ro-biphoton-spectral}
\hat{\rho}_{\Omega}=\text{Tr}_{\mathbf{q}}(\hat{\rho}),
\end{equation}
where the partial traces are done over the two-photon spectral and spatial degrees of freedom, respectively.

%%%%%%%%%%%%%%%%%%%%%%%%%%%%%%%%%%%%%%%%%%%%%%%%%%%%%%%%%%%%%%%
\subsection{HSP properties}

The characteristics of HSPs produced via SPDC will be determined by the SPDC mode function properties calculated in the previous section. In particular, it is interesting to study the purity and the heralding efficiency of the HSP. Regarding purity, we are interested in the spatial purity, $P_{\mathbf{q}_\mu}$, and spectral purity, $P_{\Omega_\mu}$, for each of the down converted photons. These purities can be calculated through the signal-idler spatial density matrix, $\hat{\rho}_{\mathbf{q}}$, from Eq.~(\ref{ro-biphoton-spatial}), and the signal-idler spatial density matrix, $\hat{\rho}_{\Omega}$, from Eq.~(\ref{ro-biphoton-spectral}). Explicitly, for the signal photon $\hat{\rho}_{\mathbf{q}_s}=\text{Tr}_{\mathbf{q}_i}(\hat{\rho}_\mathbf{q})$ and $\hat{\rho}_{\Omega_s}=\text{Tr}_{\Omega_i}(\hat{\rho}_\Omega)$, while for the idler photon $\hat{\rho}_{\mathbf{q}_i}=\text{Tr}_{\mathbf{q}_s}(\hat{\rho}_\mathbf{q})$ and $\hat{\rho}_{\Omega_i}=\text{Tr}_{\Omega_s}(\hat{\rho}_\Omega)$.
From these expressions, the spatial and spectral purities of the HSP can be written as
\begin{equation}
\label{eq:spatialpurities}
P_{\mathbf{q}_\mu}= \text{Tr}(\hat{\rho}_{\mathbf{q}_\mu}^2),
\end{equation}
and
\begin{equation}
\label{eq:spectralpurities}
P_{\Omega_\mu}= \text{Tr}(\hat{\rho}_{\Omega_\mu}^2),
\end{equation}
revealing that the purities of each individual photon are determined by the properties of the SPDC and the filtering.

On the other hand the heralding, efficiency is defined as the conditional probability of detecting a single photon given the detection of its twin. For example, taking the signal photon as the heralding one, and assuming an ideal detection system, the maximum heralding efficiency, $\eta_s$, when announcing with the signal, is defined as \cite{Klishko88}

\begin{equation}
\eta_s=\frac{\mathcal{P}_{c}}{\mathcal{P}_s},\label{eq:he}
\end{equation}
where
\begin{equation}
\mathcal{P}_s=\int d^2\mathbf{q}_sd^2\mathbf{q}_id\Omega_s d\Omega_i\\|\phi(\mathbf{q}_s,\Omega_s,\mathbf{q}_i,\Omega_i)f_s(\Omega_s)C_s(\mathbf{q}_s)|^2\label{eq:ps}
\end{equation}
is the probability of detecting the heralding signal photon after a set of filters $f_s(\Omega_s)C_s(\mathbf{q}_s)$, and
\begin{equation}
\mathcal{P}_{c}=\int d^2\mathbf{q}_sd^2\mathbf{q}_id\Omega_s d\Omega_i |\tilde{\phi}(\mathbf{q}_s,\Omega_s,\mathbf{q}_i,\Omega_i)|^2\label{eq:pc}
\end{equation}
is the joint probability of detecting the signal together with the corresponding heralded idler photon, after passing a set of filters $f_s(\Omega_s)C_s(\mathbf{q}_s)$ and $f_i(\omega_i)C_i(\mathbf{q}_i)$, respectively. Likewise, the idler maximum heralding efficiency is defined by $\eta_i=\mathcal{P}_{c}/\mathcal{P}_i$, with $\mathcal{P}_i$ the probability of detecting the heralding idler photon after a set of filters $f_i(\Omega_i)C_i(\mathbf{q}_i)$.

An intuitive idea of the probabilities $\mathcal{P}_{s}$, $\mathcal{P}_{i}$, and $\mathcal{P}_{c}$ can be extracted from the areas associated to the plots in Fig.~\ref{fig:effectofspectralfiltering} and Fig.~\ref{fig:effectofspatialfiltering}. The areas of Fig.~\ref{fig:effectofspectralfiltering}(b), Fig.~\ref{fig:effectofspectralfiltering}(c) and Fig.~\ref{fig:effectofspectralfiltering}(d) correspond to $\mathcal{P}_c$, $\mathcal{P}_s$ and $\mathcal{P}_i$, respectively. Analogously, Fig.~\ref{fig:effectofspatialfiltering}(b), \ref{fig:effectofspatialfiltering}(c) and \ref{fig:effectofspatialfiltering}(d) obey the same correspondence but for the spatial case.

The properties of the SPDC photons and the filtering process will determine the purity and heralding efficiency of the HSP. In what follows, we will use Eq.~(\ref{matrix}) in order to illustrate these dependences.

%%%%%%%%%%%%%%%%%%%%%%%%%%%%%%%%%%%%%%%%%%%%%%%%%%%%%%%%%%%%%%%
\section{Spatial properties of filtered HSP with and without spatial-spectral correlation} \label{Sec:Spatial}
With the theoretical model developed for the SPDC correlations and the definitions introduced in the previous section, we can proceed to calculate the spatial purity and heralding efficiency of a HSP produced by a filtered SPDC source. The effects of spectral filtering on HSP sources for the case $\mathbf{q}_s=\mathbf{q}_i=0$ have been previously considered \cite{Pittman05,Cassemiro10,Osorio13}. Here, we focus our attention on the role that spatial filters play in the HSP properties. 

\begin{figure}
  \includegraphics[width=0.45\textwidth]{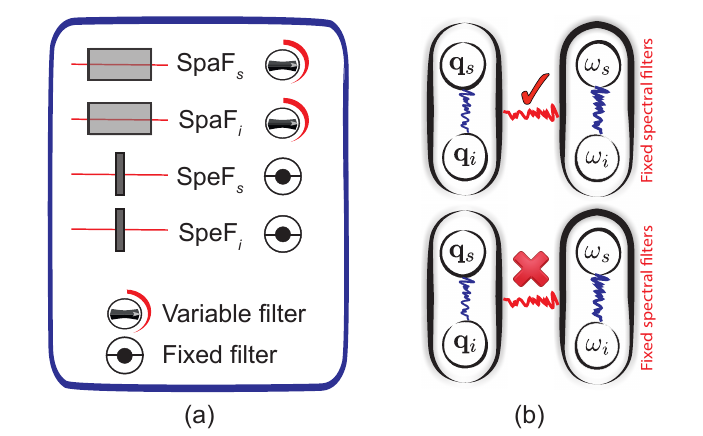}\\
  \caption{(Color online) Pictorial representation of the conditions considered to study the spatial HSP properties. (a) Scheme indicating that spatial filters vary and the spectral filters are fixed. This is symbolized through variable and fixed knobs. (b) Cartoon showing the different correlations when the spatial-spectral correlation exists (upper part), and when such correlation is completely broken (lower part.)}\label{cartoon-spatial-effects}
\end{figure}

We study the spatial purity and heralding efficiency when the HSP is either the signal or the idler photon, according to the setups of Fig.~\ref{fig:Setup}. When the signal is the heralded photon, the quantities of interest are $P_{\mathbf{q}_s}$ and $\eta_i$, whereas when the idler is the heralded photon, one is concerned with $P_{\mathbf{q}_i}$ and $\eta_s$. With a fixed value for the spectral-filter bandwidths, we vary equally the collecting mode of the spatial filters, Fig.~\ref{cartoon-spatial-effects}(a). 

\begin{figure}
  \includegraphics[width=0.45\textwidth]{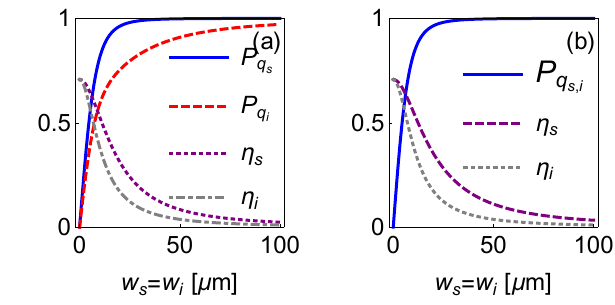}\\
  \caption{(Color online) Spatial HSP properties when the collecting modes of both photons, signal and idler, are equal and vary simultaneously. The pump properties are $\sigma_p=1.0$ nm and $\text{w}_p=10\ \mu$m. In (a), we plot purities and heralding efficiencies for the case in which the spatial-spectral correlation is present ($\sigma_s=\sigma_i=5.0$ nm.) In (b), HSP properties when the spatial-spectral correlation is broken ($\sigma_s=\sigma_i=0.1$ nm.)}\label{spatial-effects-presence-absence}
\end{figure}

We consider spectral-filter bandwidths ($\sigma_s=\sigma_i=5.0$ nm) such that the spatial-spectral correlation is not destroyed, as illustrated in the upper part of Fig.~\ref{cartoon-spatial-effects}(b). The corresponding spatial purities and heralding efficiencies are shown in Fig.~\ref{spatial-effects-presence-absence}(a). Due to the spatial-spectral correlation, the spatial subsystem is not pure and therefore we observe that $P_{\mathbf{q}_s}$ is not equal to $P_{\mathbf{q}_i}$. Additionally, $P_{\mathbf{q}_s}$ and $P_{\mathbf{q}_i}$ tend to 1 as collecting modes increase, as expected from strong filtering conditions.

Now, we consider a case in which the spatial-spectral correlation is broken, but the signal-idler correlation exist. To do so, we fix $\sigma_s=\sigma_i=0.1$ nm, as represented in the lower part of Fig.~\ref{cartoon-spatial-effects}(b). Fig.~\ref{spatial-effects-presence-absence}(b) shows how, in this situation, $P_{\mathbf{q}_s}=P_{\mathbf{q}_i}$, as expected when the spatial-spectral correlation has been destroyed.

In both graphs of Fig.~\ref{spatial-effects-presence-absence}, the heralding efficiencies $\eta_s$ and $\eta_i$ are clearly different to each other. For example, if we set 10 $\mu$m collecting modes in both paths, $\eta_s=0.55$ whereas $\eta_i=0.39$, exhibiting a difference of 29\% with respect to $\eta_s$. %This happens because the two photons have orthogonal polarizations. Hence, a filtering in the signal arm yields different results than a filter in the idler arm, as was commented for the spatial joint-spectrum of Fig.~\ref{fig:effectofspatialfiltering}. 
One can explain the differences in the heralding efficiencies by comparing the areas in Fig.~\ref{fig:effectofspatialfiltering}(c) and Fig.~\ref{fig:effectofspatialfiltering}(d), which go with $\mathcal{P}_s$ and $\mathcal{P}_i$, respectively. From these figures, one sees that $\mathcal{P}_s<\mathcal{P}_i$, indicating that a filter in the signal arm leads to a lower probability of detecting its partner than a filter in the idler arm. By means of Eq.~(\ref{eq:he}), one explains why $\eta_s>\eta_i$ in Fig.~\ref{spatial-effects-presence-absence}.

%One way to visualize the difference in heralding efficiencies is remembering the equivalence between the probabilities $\mathcal{P}_s$, $\mathcal{P}_i$ and $\mathcal{P}_c$, defined in Eq. (\ref{eq:ps}) and (\ref{eq:pc}), with the areas of Fig.~\ref{fig:effectofspatialfiltering}. Comparing Fig.~\ref{fig:effectofspatialfiltering}(c) with Fig.~\ref{fig:effectofspatialfiltering}(d), one sees that $\mathcal{P}_s<\mathcal{P}_i$, which means that a filter in the signal arm leads to a lower probability of detecting its partner than a filter in the idler arm. By means of Eq.~(\ref{eq:he}), one explains why $\eta_s>\eta_i$ in Fig.~\ref{spatial-effects-presence-absence}.

Figure~\ref{spatial-effects-presence-absence} also shows a trade-off between spatial purities and heralding efficiencies when varying the spatial filters. This is because, as the collecting modes increase, less transverse modes are coupled into the fibers, producing more pure spatial states. However, the probability of detecting the two photons from a pair reduces, resulting in a low heralding efficiency.

\begin{figure}
  \includegraphics[width=0.5\textwidth]{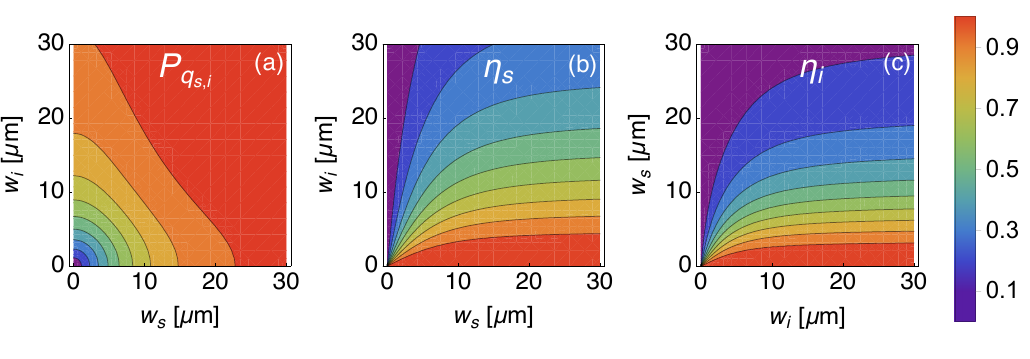}\\
  \caption{(Color online) HSP properties as a function of the signal and idler collecting modes when they vary independently. We show contour plots for the spatial HSP purity (a) and heralding efficiencies, $\eta_s$ and $\eta_i$, in (b) and (c), respectively.}\label{Spatial-HSP-Contourn}
\end{figure}

Up to now, we have considered equal collecting modes for signal and idler. Another scenario that may be implemented in the laboratory is when such collecting modes vary independently. Fig.~\ref{Spatial-HSP-Contourn}(a), Fig.~\ref{Spatial-HSP-Contourn}(b) and Fig.~\ref{Spatial-HSP-Contourn}(c) show contour plots for the spatial HSP purity and heralding efficiencies $\eta_s$ and $\eta_i$, respectively, for the case in which the spatial-spectral correlation is broken. From Fig.~\ref{Spatial-HSP-Contourn}(a), one observes that there is a set of values for the signal and idler collecting modes which leads to an almost 100\% purity. However, from a certain value of $\text{w}_s$ ($\text{w}_s\gtrsim23$ $\mu$m), the HSP purity is 100\%, no matter the collecting mode on the idler photon path. Something similar occurs for a certain value of $\text{w}_i$ ($\text{w}_i\gtrsim30$ $\mu$m). This means that one can even avoid the use of filters in one of the photon paths and still obtain a high purity. Nevertheless, these values of $\text{w}_s$ and $\text{w}_i$ are different for the signal than for the idler, giving again a difference between placing a filter on either path.

Regarding the heralding efficiencies, by looking at Fig.~\ref{Spatial-HSP-Contourn}(b) and Fig.~\ref{Spatial-HSP-Contourn}(c), we observe that a high heralding efficiency is obtained by fixing a small collecting mode for the heralded photon and increasing the heralding one. The latter can be seen in Fig.~\ref{Spatial-HSP-Contourn}(b), where $\eta_s$ approaches to 1 if one sets an idler (heralded photon) collecting mode less than 10 $\mu$m and increases the signal (heralding) collecting mode. A similar result is obtained for $\eta_i$, Fig.~\ref{Spatial-HSP-Contourn}(c), but fixing a small signal (heralded) collecting mode and varying the idler (heralding) one. Moreover, $\eta_s>\eta_i$, in agreement with Fig.~\ref{spatial-effects-presence-absence}.

We conclude that there is an asymmetry between using the signal or the idler as HSP. Since $\eta_s>\eta_i$, it is more convenient to use the ordinary polarized photon (idler photon in this work) as HSP. However, it is important to point out that this asymmetry disappears when the waist of the pump becomes wider. For the configuration we are considering, this happens for a pump waist greater than $100$ $\mu$m.

%%%%%%%%%%%%%%%%%%%%%%%%%%%%%%%%%%%%%%%%%%%%%%%%%%%%%%%%%%%%%%%
\section{Spectral properties of filtered HSP with and without spatial- spectral correlation}\label{Sec:Spectral}

\begin{figure}[ht]
  \includegraphics[width=0.45\textwidth]{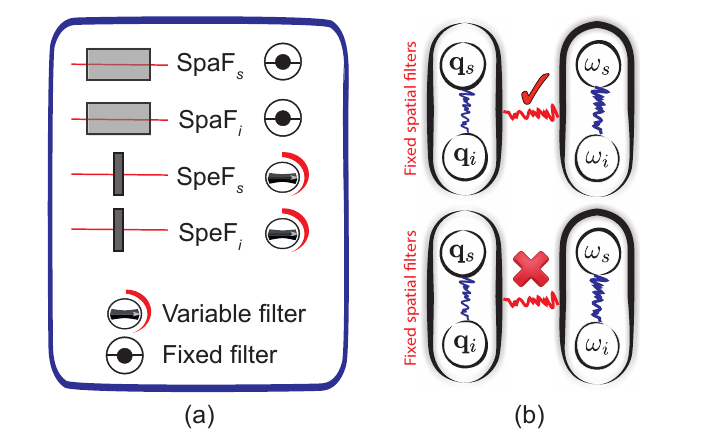}\\
  \caption{(Color online) Pictorial representation of the conditions considered to study the spectral HSP properties. (a) Scheme indicating that spatial filters are fixed and the spectral filters vary. (b) Cartoon showing the different correlations when the spatial-spectral correlation exists (upper part), and when such correlation is completely broken (lower part.)}\label{cartoon-spectral properties}
\end{figure}

In this section, we compare spatial and spectral purities, and also study the conditions to obtain photons with both high purity and high heralding efficiency. We perform this analysis of the HSP properties when the spatial filters are fixed and the spectral filters are allowed to vary equally. The values of the spatial filters are chosen in such a way that we are able to study the HSP properties in the presence and absence of the spatial-spectral correlation, as in the previous section. These conditions are illustrate in Fig.~\ref{cartoon-spectral properties}.

\begin{figure}[ht]
  \includegraphics[width=0.45\textwidth]{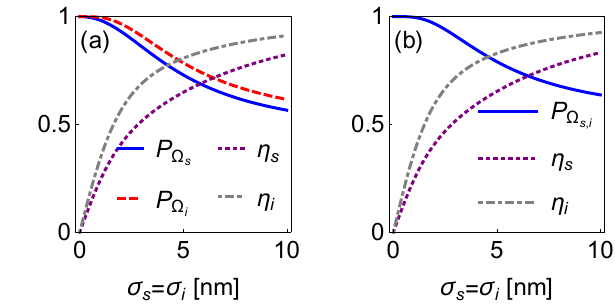}\\
  \caption{(Color online) Spectral HSP properties when the spectral-filter bandwidths of both photons, signal and idler, are equal and vary simultaneously.  The pump properties are $\sigma_p=1.0$ nm and $\text{w}_p=10\ \mu$m. In (a), we plot purities and heralding efficiencies for the case in which the spatial-spectral correlation is present ($\text{w}_s=\text{w}_i=50$ $\mu$m.) In (b), HSP properties when the spatial-spectral correlation is broken ($\text{w}_s=\text{w}_i=100$ $\mu$m.)}\label{spectral-HSP-Properties}
\end{figure}

Figure~\ref{spectral-HSP-Properties} shows spectral purities for signal and idler, $P_{\mathbf{\Omega}_s}$ and $P_{\mathbf{\Omega}_i}$, as well as heralding efficiencies, $\eta_s$ and $\eta_i$. The behavior of these HSP properties can be understood analogously to the spatial case, with the difference that a wider spectral-filter bandwidth means not filtering, while in the spatial case a greater collecting mode implies a stronger filter. Hence, contrary to Fig. \ref{spatial-effects-presence-absence}, the spectral purities become smaller and the heralding efficiencies increase as a function of $\sigma_s$ and $\sigma_i$. Concerning the heralding efficiencies, the asymmetry between announcing with the signal or idler is still present, but now $\eta_i>\eta_s$ in contrast to the spatial case. Comparing Fig.~\ref{fig:effectofspectralfiltering}(c) with Fig.~\ref{fig:effectofspectralfiltering}(d), this fact is explained if one realices that, contrary to the spatial case, in the frequency domain $\mathcal{P}_s>\mathcal{P}_i$, leading to $\eta_s<\eta_i$.

So far, we have seen how is the behavior of the spectral purity and heralding efficiency separately. In what follows, we will concentrate in which conditions must be accomplished to get photons with both high spectral purity and high heralding efficiency. In order to find these conditions, we define the spectral purity-efficiency factors (PEFs) as $P_{\Omega_s}\eta_i$ when the signal photon is the HSP and as $P_{\Omega_i}\eta_s$ when the idler photon is the heralded one. PEFs satisfy
\begin{subequations}
\begin{align}\label{eq:purities-efficiencies}
0\leq& P_{\Omega_s}\eta_i\leq 1,\\
0\leq& P_{\Omega_i}\eta_s\leq 1,
\end{align}
\end{subequations}
displaying their maximum values when both the spectral purity and the corresponding heralding efficiency reach independently their highest possible values.

\begin{figure}[htb]
  \includegraphics[width=0.45\textwidth]{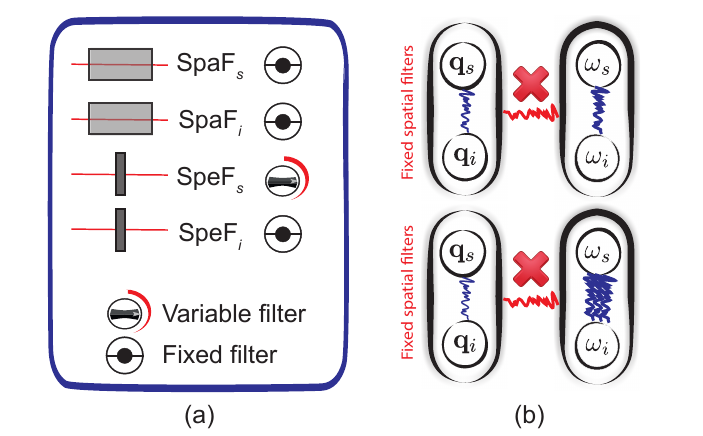}\\
  \caption{(Color online) Pictorial representation of the conditions considered to study the behavior of PEFs. (a) Scheme indicating that spatial filters are fixed and only one of the spectral filters varies. (b) Cartoon showing two different strengths for the spectral signal-idler correlation. The multiple lines in the lower part indicates that the spectral correlation is stronger than in the upper part.}\label{cartoon-PEF}
\end{figure}

We consider the behavior of PEFs for the filtering configuration illustrated in Fig.~\ref{cartoon-PEF}(a), in which the spatial filters are fixed and only one of the spectral filters varies. We concentrate in the case in which we have the spatial-spectral correlation together with the spectral signal-idler correlation. This is schematically represented in the upper part of Fig.~\ref{cartoon-PEF}(b).

In Fig. 11(a), we plot PEF when the idler is the HSP, according to the setup of Fig. 1(a), for two values of the idler's spectral filter. For a weak filter in the idler's path ($\sigma_i=10$ nm), the maximum PEF is obtained by making a strong filtering in the signal (heralding) photon ($\sigma_s<1$ nm), while for a strong filtered idler photon ($\sigma_i=0.1$ nm) PEF is almost zero, no matter the filter in the signal arm. Fig. 11(b) shows PEF when the signal is the HSP, corresponding to the setup of Fig. 1(b). We compare again two values of the spectral filter for the idler photon, but now this photon plays the role of the heralding one. In both cases, PEF increases when the filter in the signal arm increases. However, the maximum PEF is achieved when the filter of the idler (heralding) photon becomes narrow ($\sigma_i=0.1$ nm).
%practically without filtering the idler (heralding) photon ($\sigma_i=10$ nm).

\begin{figure}[htb]
  \includegraphics[width=0.45\textwidth]{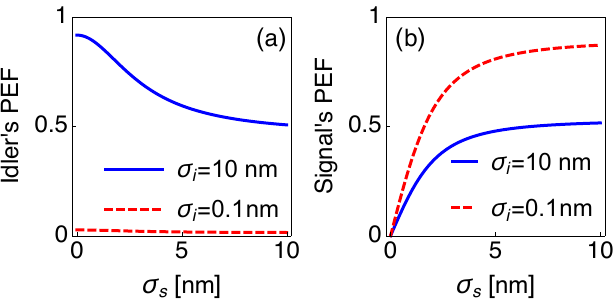}\\
  \caption{(Color online) PEF for a specific value of the spectral correlation between signal and idler. The pump properties are $\sigma_p=1.0$ nm and $\text{w}_p=10\ \mu$m, and the spatial-spectral correlation is present making $\text{w}_s=\text{w}_i=50$ $\mu$m. In (a), we plot idler's PEF when this photon is the HSP, while in (b) the PEF is shown when the signal is the HSP.}\label{PEF-weak}
\end{figure}

\begin{figure}[htb]
  \includegraphics[width=0.45\textwidth]{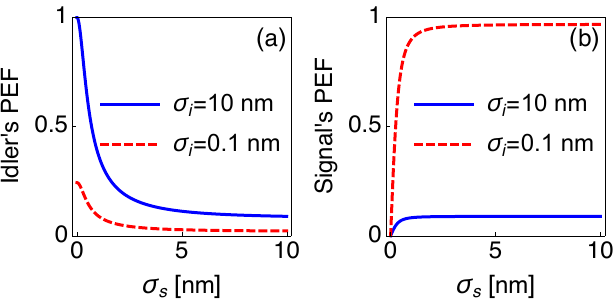}\\
  \caption{(Color online) PEF when the spectral signal-idler spectral correlation is stronger than in Fig. \ref{PEF-weak}. The pump properties are $\sigma_p=0.1$ nm and $\text{w}_p=10\ \mu$m, and the spatial-spectral correlation is present making $\text{w}_s=\text{w}_i=50$ $\mu$m. In (a), we plot idler's PEF when this photon is the HSP, while in (b) the PEF is shown when the signal is the HSP.}\label{PEF-strong}
\end{figure}

The lower part of Fig. \ref{cartoon-PEF}(b) shows the effect of increasing the spectral signal-idler correlation, which can be accomplish by, for example, decreasing the bandwidth of the pump. In comparison with plots in Fig.~\ref{PEF-weak}, a similar behavior is observed when the HSP is the idler, Fig.~\ref{PEF-strong}(a), or the signal, Fig.~\ref{PEF-strong}(b). However, when the spectral signal-idler correlation is stronger, the variations of the PEF are steeper, restricting the implementation of the idler photon as HSP to the use of a very narrow filter in the signal arm ($\sigma_s\ll 1$ nm and $\sigma_i= 10$ nm), as can be seen in Fig.~\ref{PEF-strong}(a). On the other hand, when the signal is the HSP, a 100\% PEF is obtained when the idler (heralding) filter is $\sigma_i=0.1$ nm and the signal photon is almost unfiltered, according to Fig.~\ref{PEF-strong}(b). This result reveals that one can have a spectrally pure HSP source with a high heralding efficiency through the use of filters.

%%%%%%%%%%%%%%%%%%%%%%%%%%%%%%%%%%%%%%%%%%%%%%%%%%%%%%%%%%%%%%%
\section{Conclusions}\label{Sec:Conclusions}
We described the spatial and spectral purities as well as heralding efficiencies of HSP produced via collinear type-II SPDC, considering the effects of spatial and spectral filters. As a new result, we found that the heralding efficiencies depend on the photon that is used to announce the presence of its partner. In particular, when dealing with the spatial properties of the single photon, the highest heralding efficiency was obtained when the ordinary-polarized photon (idler) was used as the HSP. On the other hand, when one is concerned about the spectral properties of the single photon, it is more convenient to implement the extraordinary-polarized photon (signal) as the HSP. Additionally, we observed the expected behavior for the spatial and spectral purity of signal and idler depending on the presence of the spatial-spectral correlation.

Finally, we introduce the purity-efficiency factor (PEF) to figure out which conditions must be satisfy in order to build a suitable HSP source based on SPDC. We found that in the spectral case it is possible to have high purity and high heralding efficiency simultaneously. This is relevant since new sources that produce high number of photon pairs have been develop in recent years, and the use of filters in an experimental setup is more convenient than other approaches to control correlations.

%%%%%%%%%%%%%%%%%%%%%%%%%%%%%%%%%%%%%%%%%%%%%%%%%%%%%%%%%%%%%%%
\begin{acknowledgments}
O. C. and A.V. acknowledge support from Faculty of Science and Vicerector\'ia de Investigaciones of Universidad de los Andes, Bogot\'a, Colombia. C. I. O. work is supported by NanoNextNL, a micro- and nanotechnology consortium of the Government of The Netherlands and 130 partners.
\end{acknowledgments}

\bibliography{References}

\end{document}